\documentclass[reprint,onecolumn, tightenlines,nofootinbib,
notitlepage,longbibliography]{revtex4-1}

\usepackage{graphicx} 
\usepackage{float}
\usepackage{array} 
\usepackage{paralist} 
\usepackage{subfig}
\usepackage{fancyhdr} 
\usepackage{qcircuit}
\usepackage{amsmath,amsthm}
\usepackage{mathtools}
\usepackage{enumitem}
\usepackage{xcolor}
\newtheorem{rul}{Rule}{\bfseries}{\itshape}
\newcommand{\mH}{\mathcal{H}}
\newcommand{\mU}{\mathcal{U}}
\newcommand{\bU}{\bar{U}}
\newcommand{\ket}[1]{\ensuremath{\left|#1\right\rangle}} 
\newcommand{\bra}[1]{\ensuremath{\left\langle#1\right|}} 

\renewcommand{\bf}[1]{\ensuremath{\mathbf{#1}}}
\begin{document}
\title{A Generalized Circuit  for the Hamiltonian Dynamics Through the Truncated  Series
} 
 \author{Ammar~Daskin}
\affiliation{Department of Computer Engineering, Istanbul Medeniyet University, Uskudar, Istanbul, Turkey}
\author{Sabre~Kais}
\affiliation{Department of Chemistry, Department of Physics and Birck Nanotechnology Center, Purdue University, West Lafayette, IN, USA}
\begin{abstract}
In this paper, we present a method for the Hamiltonian simulation in the context of eigenvalue estimation problems which improves earlier results dealing with Hamiltonian simulation through the truncated Taylor series.
In particular, we present a fixed-quantum circuit design for the simulation of the Hamiltonian dynamics, $\mH(t)$, through the truncated Taylor series method described by Berry et al. \cite{berry2015simulating}.
The circuit is general and can be used to simulate any given matrix in the phase estimation algorithm by only changing the angle values of the quantum gates implementing the time variable $t$ in the series.
The circuit complexity depends on the number of summation terms composing the Hamiltonian and requires $O(Ln)$ number of quantum gates for the simulation of a molecular Hamiltonian. Here, $n$ is the number of states of a spin orbital, and $L$  is the number of terms in the molecular Hamiltonian and generally bounded by $O(n^4)$.   
We also discuss how to use the circuit in adaptive processes and eigenvalue related problems along with a slight modified version of the iterative phase estimation algorithm.
In addition,  a simple divide and conquer method is presented for mapping a matrix which are not given as sums of unitary matrices into the circuit. The complexity of the circuit is directly related to the structure of the matrix and can be bounded by $O(poly(n))$ for a matrix with $poly(n)-$sparsity. 
\end{abstract}

\maketitle
\section{Introduction}
Quantum phase estimation \cite{Kitaev1996} is a computationally powerful algorithm used in the study of various eigenvalue problems. 
It is the key component of quantum chemistry simulations 
\cite{somma2002simulating,brown2010using,kassal2011simulating,wecker2014gate} and many other quantum algorithms (see the recent review article 
\cite{montanaro2016quantum} or the book \cite{nielsen2002quantum}) such as the Shor's integer factorization \cite{shor1999polynomial} and HHL algorithm for the linear systems of equations \cite{harrow2009quantum}. 
Given a unitary matrix $U$ with an approximate eigenvector $\ket{\varphi}$,
since any eigenvalue of a unitary matrix is in the form of a complex exponential $e^{i2\pi\phi}$ for $0 \leq \phi< 1$, 
the algorithm particularly estimates the value of $\phi$.
In quantum simulations, since $U$ is the time evolution operator of a Hamiltonian $\mH$ representing the dynamic of a quantum system, i.e. $U = e^{i\mH t}$, the estimated value  also yields an eigenvalue of $\mH$. 
Therefore, the algorithm is used to find the eigenvalues (generally the lowest corresponding to the ground state energy) of $\mH$. 

 Simulating $\mH$ of a quantum system  through the phase estimation algorithm necessitates  an explicit circuit design  of $e^{i\mH t}$ in terms of quantum gates.
For a given $\mH  = \sum^{L-1}_{l=0} H_l$,  the generalized Trotter formula \cite{trotter1959product,suzuki1976generalized} is a common way to estimate $e^{i\mH t}$ as a product of evolution operators $e^{iH_lt}$ which can be mapped to quantum gates. 
The resulting product accurately yields the evolution if all terms in the formula commute with each other. 
Otherwise, it involves an error  which depends on the order of the approximation.
The amount of the error and the computational complexity (the required number of quantum gates) also increase proportionally with the number of terms $L$ , $|| Ht ||$, and the simulation accuracy \cite{poulin2015trotter}.
 
 In quantum computing, the complexity of implementing a circuit for the Hamiltonian can be decreased by using additional subspaces.
Berry et al.\cite{berry2015simulating} have proposed using the Taylor expansion of $e^{i\mH t}$ directly on quantum circuits by adding an ancillary register to the system. 
In Ref.\cite{daskin2017ancilla}, for a Hermitian $\mH$ we have showed that when  $\sqrt{I-\mH ^2}$ is available,  one can use the following unitary matrix in the phase estimation algorithm:
\begin{equation}
\label{EqHsqrtH}
    \left(\begin{matrix}
    \mH   & -\sqrt{I-\mH ^2}\\
     \sqrt{I-\mH ^2} & \mH  
    \end{matrix}\right).
\end{equation}
This notion of using an extended system to simulate a smaller one is  generalized as quantum signal processing \cite{low2017optimal}, where a unitary matrix similar to Eq.\eqref{EqHsqrtH} used without the phase estimation algorithm. 
Recently, the overhead of the truncated Taylor series method is reduced by changing the computational basis to attain the square root of a  matrix efficiently \cite{poulin2017fast}. 
In  Ref.\cite{daskin2017direct}, successive applications of $\left(I+i\mH /\mu\right)$ with $\mu\geq 10||\mH ||$ are used to obtain the eigenvalue of $\mH $ from the sine value of the phase in the phase estimation algorithm. Here, note that these approaches are assumed that the Hamiltonian is given as a sum of simple unitary matrices. 
In Ref.\cite{babbush2018exponentially}, the truncated Taylor series method is also used for quantum simulations after decomposing the configuration interaction matrix into a sum of sparse matrices. 

The main contribution of this paper is as follows:
\begin{itemize}
\item  Given a Hamiltonian $\mH$ we consider 
$U ( t ) = \left( t \mH + i\sqrt{ I - t^2 \mH^2} \right)$ and approximate
the expression involving the square root by $\left(I - t^2 \mH^2/2\right)$ to obtain a circuit simulating the Hamiltonian dynamics with reduced complexity. 
Using the circuit in the phase estimation algorithm, the eigenvalue of $\mH$ can be obtained from the cosine value of the phase. 
The presented circuit has a fixed design and can be used in adaptive processes along with a modified iterative phase estimation algorithm.  
\item We also describe a divide and conquer method that can be used to write a general matrix as a sum of unitary matrices. 
The method groups matrix elements into submatrices and directly maps them to the quantum gates. 
Because of this direct mapping, the number of quantum gates is related to the number of nonzero matrix elements and can be reduced in the case of structured-sparse matrices. 
\end{itemize}

The remaining part of this paper is organized as follows: 
In the following subsection, we summarize the truncated Taylor series method. 
In the next section; we first describe the Taylor expansion used in this paper, then present a general circuit design for the described expansion, then analyze its complexity, then explain how to use it in the phase estimation algorithm, and then discuss the molecular Hamiltonians and the Hamiltonian for the hydrogen molecule as example system.
In Sec.\ref{SecIII}, we explain the divide and conquer method and analyze the complexity in terms of the required number of CNOTs.
In Sec.\ref{Sec4}, we discuss how the method can be used with structured matrices and use the Hamiltonian of the hydrogen molecule as an example system. We also discuss adaptive processes and describe a modification to the iterative phase estimation algorithm.
In the final section, we conclude the paper.

\subsection{Truncated Taylor Series Method}
For a given matrix $\bU$, assume that we are able to build the circuit equivalent of the following unitary matrix by using an ancilla quantum register:
\begin{equation}
\label{EqUbig}
\mU=\left(\begin{matrix}
\bU & \bullet \\ \bullet& \bullet
\end{matrix}\right),
\end{equation}
where each ``$\bullet$" represents a matrix which has no special meaning and their dimensions may be different on the diagonal and anti-diagonal of the matrix.
When applied to any arbitrary $\ket{\psi}$ on the system register, the above matrix  generates the following output state: 
\begin{equation}
\label{EqGeneralState}
\mU\ket{\bf 0}\ket{\psi} = \ket{\Phi} + \ket{\bf0}\bU\ket{\psi}.
\end{equation}
Here, \ket{\bf 0} represents the first vector in the standard basis and \ket{\Phi} is the part of the output in which the first register is not in \ket{\bf0} state. 
In this output, when the first register is in \ket{\bf0} state,  the second register holds $\bU\ket{\psi}$.
Therefore, $\mU$ can be used to emulate the action of $\bU$ on any arbitrary state \ket{\psi}.
This idea is used in various contexts: e.g., in Ref.\cite{daskin2012universal}, a programmable circuit design is presented for unitary matrices.
Given a Hamiltonian $\mH  = \sum^{L}_{l=1} \alpha_lH_l$ with $H_l$ representing a unitary matrix; 
the Taylor expansion of $	e^{i\mH t} $ truncated at the $K$th order is defined as:
\begin{equation}
\label{EqExpansionProduct}
	U\left(t\right) = e^{i\mH t} \approx	\bU\left(t\right) 
    = \sum_{k=0}^{K} 
    \frac{\left(i\mH t\right)^k}{k!}.
\end{equation}
As in Ref.\cite{berry2015simulating}, we obtain:
\begin{equation}
\label{EqExpansionProduct}
	\bU\left(t\right) 
    = \sum_{k=0}^{K} \sum_{l_1,\dots l_k=0}^{L-1} 
    \frac{(it)^k}{k!}\alpha_{l_1}\dots \alpha_{l_k}H_{l_1}\dots H_{l_k} 
    = 
    \sum_{j=0}^{M-1} \beta_j V_j,
\end{equation}
Here, while $V_j$s are some products of $H_l$s, $\beta_j$s are of $\alpha_l$s. And $M \approx L^K$ is the number of terms.
The above expansion can be implemented as a circuit by using the following \cite{berry2015simulating}:
\begin{equation}
\label{EqH1}
\mathcal{U} = \left(B^*\otimes I\right)V\left(B\otimes I\right) = \left(\begin{matrix}
\bU & \bullet \\ \bullet& \bullet
\end{matrix}\right).
\end{equation}
where $V = blkdiag\left(V_0, V_1, \dots, V_M\right)$ and $B = \sum_{j=0}^{M-1}\sqrt{\beta_j} \ket{j}$.  
$V$ can be implemented as a circuit by using an additional $\approx logM$ control qubits: When these qubits are in the state \ket{\bf j},  $V_j$ is applied to the system register. 

\section{Expansion Formula and the Circuit}
Given a Hamiltonian 
$\mH\in\bf{R}^N$ with  $\left(N=2^n\right)$;
when $||tH||\leq 1$ (see the footnote\footnote{If $\mH$ is given as a sum of unitaries, then normalizing the coefficients directly make $||H||\leq 1$. If it is given as a matrix, then one can divide the matrix elements by the 1- or inifinity-norm of the matrix, which can be computed in $poly(n)$ time if there are $poly(n)$ number of nonzero matrix elements.  }),  as in  Ref.\cite{babbush2018exponentially, babbush2017improved, low2017hamiltonian, low2016hamiltonian}) we define: 
\begin{equation}
\label{EqUtSquareRoot}
U \left(t\right) = t\mH  + i\sqrt{I - t^2\mH ^2}.
\end{equation}
Here, $U\left(t\right)$ describes a unitary matrix \cite{wu1994additive}
with the eigenvalues whose real parts are equal to those of the Hamiltonian. In addition, $U\left(t\right)$  and $t\mH $ have the same eigenvectors.
Truncating the Taylor expansion of the square root at the second term, we obtain the following approximation:
\begin{equation}
\label{EqFinalUt}
 \bU \left(t\right)= t\mH  + i\left(I - \frac{t^2\mH ^2}{2}\right),
\end{equation}
which is the same as the Taylor expansion of $i e^{-i\mH t}$ truncated at the third term.
The most significant part of this approximation is that it does not introduce any error on the real part of the eigenvalues of $U\left(t\right)$. 
Hence, by mapping $\bU \left(t\right)$ to a circuit and using the phase estimation algorithm, one can obtain the eigenvalue from the cosine value of the phase.

In the following subsection, based on Eq.\eqref{EqH1}, a circuit design is presented for $\bU\left(t\right)$.
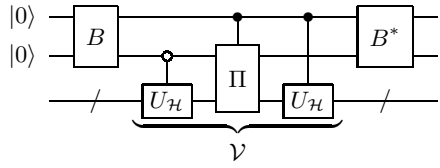
\begin{figure}
\begin{center}
\mbox{
\Qcircuit @C=1em @R=.7em {
\lstick{\ket{0}} & \multigate{1}{B} & \qw            & \ctrl{1}          & \ctrl{2} & \multigate{1}{B^*}& \qw \\
\lstick{\ket{0}} & \ghost{B}		   & \ctrlo{1} 	   & \multigate{1}{\Pi}& \qw       & \ghost{B^*} & \qw & \\
    				& {/} \qw 		   & \gate{U_\mH}   & \ghost{\Pi}       &  \gate{U_\mH}&\qw{/}& \qw 
    				\gategroup{1}{3}{3}{5}{.7em}{_\}}\\ \\
    				&&&\mathcal{V}&
}
}
\end{center}
    \caption{\label{FigCircuitAll}The circuit for $\mU\left(t\right)$: In total 
    $(n+logL+2)$ qubits are employed in the circuit. 
    }
\end{figure}
 \subsection{Circuit Design for $\bU\left(t\right)$}
{When $\bU\left(t\right)$ is close to a unitary matrix, one may  consider finding a circuit design through matrix decomposition techniques such as QR iterations \cite{mottonen2004circuits} or Householder transformations \cite{bullock2005asymptotically,urias2010householder}. However, such a task would require  finding the square of the Hamiltonian and be equivalent to the diagonalization of the Hamiltonian in terms of the complexity: i.e. $O(N^3)$ for an $N$ dimensional dense matrix.}
 
In this paper, we will first assume that we know how to obtain the circuit for $\mH$ in the following form:
     \begin{equation}
     \label{EqUh}
     U_\mH = \left(\begin{matrix}
     \mH & \bullet \\ \bullet& \bullet
     \end{matrix}\right).
     \end{equation}
{Here, if $\mH$ is an orthogonal matrix, then $U_\mH=\mH$. Otherwise, a ``$\bullet$" represents a matrix:} if $\mH$ is a sum of $L$ number of unitary matrices, then $U_\mH$ is  a matrix of dimension $LN\times LN$.

In Fig.\ref{FigCircuitAll}, using $U_\mH$ we draw a circuit which can emulate the action of $\bU \left(t\right)$ in the same way as shown in Eq.\eqref{EqGeneralState}.
The circuit  can be considered as the matrix product $\mU(t)=\left(B^*\otimes I\right)\mathcal{V}\left(B\otimes I\right)$, where $B$ is the coefficient matrix and $\mathcal{V}$ is the selection matrix for the terms included in $\bU \left(t\right)$:
\begin{itemize}
\item In matrix form, the gate $B$ is a $4\times4$ matrix that includes the square root of the coefficients of the expansion in Eq.\eqref{EqFinalUt}: 
\begin{equation}
B = \frac{1}{||\ket{\bf{b}}||} \left(\begin{matrix}
\sqrt{t} & 1 & t/\sqrt{2} & 0\\
1 & -\sqrt{t} &0 &t/\sqrt{2}\\
t/\sqrt{2} & 0 & -\sqrt{t} & -1\\
0 & -t/\sqrt{2} & 1 & -\sqrt{t}\\
\end{matrix} \right).
\end{equation}
Here, $\bra{\bf{b}} = \left[\sqrt{t},  1,  t/\sqrt{2},  0\right] $.
Since this matrix describes a Householder transformation, 
it can be implemented by using 4 quantum gates \cite{bullock2005asymptotically,urias2010householder}. 
\item $\mathcal{V}$ is the product of two controlled-$U_\mH$ and a controlled-permutation($\Pi$) gates. This has the following matrix form:
\begin{equation}
\mathcal{V} = \left(\begin{matrix}
 \begin{matrix}
     \mH & \bullet \\ \bullet& \bullet
     \end{matrix}& & &\\
   & I_{2N} & &\\
   & &\begin{matrix}
     \mH^2 & \bullet & \bullet& \bullet\\
     \bullet & \bullet& \bullet& \bullet\\
     \bullet & \bullet& \bullet& \bullet\\
     \bullet & \bullet& \bullet& \bullet\\
     \end{matrix}\\
  \end{matrix}\right).
\end{equation}
Because of the zero coefficient in the matrix $B$, the last part on the diagonal of this matrix are disregarded.
The construction of $\mathcal{V}$ is achieved through a permutation matrix
$\Pi$ (also used in Ref.\cite{daskin2017direct}) which realigns matrix elements to obtain $\mH^2$ on the diagonal and is defined as:
\begin{equation}
\Pi = \left(\begin{matrix}
I_N & &\\ & X \otimes I_{LN-N}&\\
&&I_N
\end{matrix} \right).
\end{equation}
If we apply this to the first controlled gate, then we attain the following product which leads to $\mathcal{V}$ given above:
\begin{align}
 \scriptscriptstyle
\left(
\begin{matrix}
  I_{2N}& & &\\
   & I_{2N} & &\\
    & &\begin{matrix}
     \mH & \bullet \\ \bullet& \bullet
     \end{matrix} &\\
   &  & &U_\mH\\
   \end{matrix}
\right)
\left(
\begin{matrix}
  U_\mH& & &\\
   & I_{2N} & &\\
    & &\begin{matrix}
     \mH & \bullet & \\ &  & I_N &\\ \bullet& \bullet
&     \end{matrix} &\\
   &  & &I_{N}\\
   \end{matrix}
\right)
\end{align}\end{itemize}
\subsection{Complexity Analysis} 
Computational complexity of a quantum circuit is generally determined by the required number of CNOT gates. Since the coefficient matrix $B$ can be implemented by 4 quantum gates, we will essentially count the number of CNOTs needed for the controlled-$U_\mH$ operations in order to estimate the computational complexity of the circuit $\mU(t)$ in Fig.\ref{FigCircuitAll} at time $t$.

{As in Ref.\cite{berry2015simulating}, let us assume that for a given $\mH  = \sum^{L-1}_{l=0} \alpha_lH_l$,  we have an oracle $select(\mH)$ that applies $H_l$ to the system register when the ancilla register in \ket{\bf l} state. Since $0\leq l\leq L-1$, $\log{L}$ number of qubits is needed in the ancilla.
 Using this selection operator along with the operator $B_\mH$ whose leading row and column are the coefficient vector $[\alpha_0, \dots, \alpha_{L-1}]$, we can assume that we have a mechanism to construct $U_\mH$ given in Eq.\eqref{EqUh}: 
 $U_\mH = (B^*_\mH \otimes I^{\otimes n}) select(H)(B_\mH \otimes I^{\otimes n})$.
 Since there are two controlled-$U_\mH$ gates in the circuit $\mU(t)$, in total $\mU(t)$ makes  $2L$ number of queries to $select(\mH)$. 
In addition, $B_\mH$ is an operator on log$L$ qubits and can be implemented by using $O(L)$ number of quantum gates as a Householder transformation. Therefore, the total complexity for the circuit can be bounded by $O(L)$.

Ref.\cite{berry2015simulating} is concerned with the Hamiltonian simulation with error $\epsilon$. The evolution, $e^{iHt}$, is divided into $r$ segments. Then each segment is approximated through the Taylor series truncated at order $K$. 
The query complexity is shown  to be proportional to $L^K$, where $K$ is set to be $O\left(\frac{\log(r/\epsilon)}{\log\log(r/\epsilon)}\right)$ to obtain the accuracy $(\epsilon/r)$ for each segment \cite{berry2015simulating,novo2016improved}. 
In contrast, in this paper the simulation is performed in the context of eigenvalue estimation 
and the error introduced by the truncation of the Taylor series at
the third term does not affect the eigenvalue error in Eq.\eqref{EqUtSquareRoot} and Eq.\eqref{EqFinalUt} (The next subsection explains how to use these equations in the phase estimation algorithm.).
Therefore, considering Eq.\eqref{EqFinalUt} in combination with Eq.\eqref{EqExpansionProduct} it is immediate
to see that one can set $K = 2$ to get a computational cost proportional to $L^2$ (as opposed to the value of $K$ in Ref.\cite{berry2015simulating} that increases with $\epsilon^{-1}$ ) which can be further reduced to $O(L)$ as done in this paper.

Note that the query complexity $O(L)$ when $L = O(poly(n))$ may appear to be small but when we
take into account the implementation cost of the queries the gate complexity may be much higher as illustrated in Ref.\cite{babbush2016exponentially} for the simulation of Electronic Hamiltonians. 
In order to have a $O(poly(n))$ total gate cost, in addition to $L = O(poly(n))$,  the required number of quantum gates for each $H_l$ should be bounded by $O(poly(n))$. 
This is the case when each $H_l$ is a  matrix with $poly(n)$-sparsity (the number of non-zero elements is bounded by $O(poly(n))$)\cite{berry2007efficient,childs2010simulating} or  an $n-$fold tensor product of Pauli matrices.

\subsection{Simulating with the Phase Estimation Algorithm}

 The phase estimation algorithm (PEA) \cite{Kitaev1996} uses  $U(t)^{2^{j-1}}$ to compute the $j$th bit of the eigenvalue $U(t)$. 
 The powers of $\tilde{U}(t)$ in Eq.\eqref{EqFinalUt} can be obtained  by successively applying the circuit $\mU(t)$ along with a projector operator. 
Since this requires $2^j$ repetitions for the $2^j$th power of $\tilde{U}(t)$, in the phase estimation algorithm $\mU(t)$ is applied  $O(2^{m})$ times to obtain an eigenvalue with $m$-bit precision.
Therefore, if the circuit $\tilde{U}(t)$ is of $O(Lpoly(n))$ number of quantum gates, then the eigenvalue can be obtained in $O(2^{m}Lpoly(n))$ time complexity.

 If 
$\tilde{U} \left(t\right)$ is considered as an approximation to $e^{i\mH t}$, then the $j$th bit can be estimated through the multiplication of $t$ by $2^j$: i.e., $\tilde{U} \left(t2^j\right) = i2^jt\mH  -\left(I - 2^{2j-1}t^2\mH^2\right)$ which is an approximation to $e^{i\mH t2^j}$. However, in this case the approximation error is  $O(t^3||H^3||)$. In addition, changing the elements of the coefficient vector  
$ \bra{\bf{b}} = \left[\sqrt{t2^j},  1,  t2^j/\sqrt{2},  0\right]$  
changes the value of $||\bra{\bf{b}}||_2$. Since $||\bra{\bf{b}}||_2$ impacts the success probability in the output, this may further affect the accuracy of the obtained eigenvalue. 
Note that $U \left(t\right)$ in Eq.\eqref{EqUtSquareRoot} (or Eq.\eqref{EqFinalUt} as an approximation to  Eq.\eqref{EqUtSquareRoot}) cannot be used in the phase estimation algorithm by simply changing the elements of $\bra{\bf{b}}$ .
Because if the eigenvalue of  $U \left(t\right)$ is $e^{i\phi}$, the eigenvalue of $U \left(2^jt\right)$ is not necessarily  $e^{i2^j\phi}$.

 \subsubsection{Example Hamiltonian}
 \label{SecIIExampleHamiltonian}
 As an example let us consider the molecular electronic Hamiltonian in the formalism of the second quantization\cite{Lanyon2010towardsH2,whitfield2011simulation,Seeley2012}:
 \begin{equation}
 \mH = \sum_{p,q}h_{p,q}a_p^\dagger a_q + \frac{1}{2}\sum_{p,q,r,s}h_{p,q,r,s}h_{p,q,r,s}a_p^\dagger a_q^\dagger a_r a_s,
 \end{equation}
where $a_j$ and $a_j^\dagger$ are the spinless fermionic creation and annihilation operators that are used to define the interaction  of  a  fermionic  system. In the context of quantum chemistry, here, $j \in \{0,...,n-1\}$ and represents the state of a spin orbital. $h_{p,q}$ and $h_{p,q,r,s}$ are one and two electron integrals classically computed through Hartree-Fock method.

In the occupation number basis, the creation and annihilation operators can be written in terms of Pauli matrices ($\sigma_x$, $\sigma_y$, $\sigma_z$) by using the Jordan-Wigner transformation\cite{Jordan1928}:
\begin{equation}
a_j \rightarrow I^{\otimes n-j-1} \otimes \sigma_+ \otimes \sigma_z^{\otimes j}, \text{\ and\ } 
a_j^\dagger \rightarrow I^{\otimes n-j-1}\otimes \sigma_-\otimes \sigma_z^{\otimes j},
\end{equation}  
where
\begin{equation}
\sigma_+ = \ket{1}\bra{0} = \frac{\sigma_x-i\sigma_y}{2}, \text{\ and\ } 
\sigma_- = \ket{0}\bra{1} = \frac{\sigma_x+i\sigma_y}{2}. 
\end{equation}  
 Alternative to the occupation number basis, the parity basis and the Bravyi-Kitaev basis \cite{BRAVYI2002210} can be used to map this Hamiltonian into the Pauli matrices (see Ref.\cite{Seeley2012} for a comparison).   
 As a particular example, we will use the hydrogen molecule in minimal basis. 
The Hamiltonian for the hydrogen molecule is given as a sum of products of Pauli matrices through Bravyi-Kitaev transformation in Eq.(79) of Ref.\cite{Seeley2012}:
\begin{equation}
\label{EqH2Hamiltonian}
\begin{split}
\mH_{H_2} = &
 -0.81261 I + 0.171201 \sigma_0^z 
+ 0.16862325 \sigma_1^z 
-0.2227965 \sigma_2^z 
+0.171201 \sigma_1^z \sigma_0^z 
+0.12054625 \sigma_2^z \sigma_0^z 
\\
&+ 0.17434925 \sigma_3^z \sigma_1^z
+0.04532175 \sigma_2^x \sigma_1^z \sigma_0^x
+0.04532175\sigma_2^y \sigma_1^z \sigma_0^y
+0.165868 \sigma_2^z \sigma_1^z \sigma_0^z
+0.12054625 \sigma_3^z \sigma_2^z \sigma_0^z
\\
&-0.2227965 \sigma_3^z \sigma_2^z \sigma_1^z
+0.04532175 \sigma_3^z \sigma_2^x \sigma_1^z \sigma_0^x
+0.04532175 \sigma_3^z \sigma_2^y \sigma_1^z \sigma_0^y
+0.165868 \sigma_3^z \sigma_2^z \sigma_1^z \sigma_0^z
\end{split}
\end{equation}
The evolution operator for this Hamiltonian is computed through Trotter-Suzuki decomposition. 
In this decomposition, the exponential (and the circuit) for the each term is computed separately (note that an $n$-fold tensor product of Pauli matrices requires $(2n-1)$ CNOT gates.). Then, these circuits are combined to estimate the whole evolution operator. Commuting terms in this Hamiltonian simplify the resulting circuit. Because of these simplifications, Ref.\cite{Seeley2012} shows that the circuit simulating the single first order Trotter time-step of this Hamiltonian requires only 44 CNOT and 30 single gates. 

 In our case, since an $n$-fold tensor product of Pauli matrices requires $n$ single gates, each term can be implemented by using only single gates. 
Since there are L = 15 terms, we need at least 4 qubits in the ancilla to control which term to be applied to the system. 
This is basically a $select(\mH)$ operator that applies the single gates in the $j$th term to the system qubits when the ancilla is in \ket{\bf j} state.
Since there are 4 system qubits, this leads to a 4 multi-controlled network.
Here, note that a network controlled by $n$ qubits can be implemented by $2^n$ CNOT gates by following  the decomposition given in Ref.\cite{mottonen2004circuits}.
Since there are 4 multi-controlled networks, in total $select(\mH)$ requires $4\times 2^4$ CNOT gates. 
Another $2^4$ CNOT gates is necessary for the implementation of the coefficients. 
Therefore, the circuit $U_\mH$ for the Hamiltonian will require $\approx 80$ CNOT gates in total.
Note that this number is a rough estimate, the number of CNOTs may be reduced by some optimization on the circuit.

Considering the one iteration of the phase estimation along with the circuit $\mU$; since there are two more control qubits for $U_\mH$, the number of CNOTs for each $U_\mH$ increases fourfold. 
For two $U_\mH$s controlled by two qubits, $2\times4\times80=640$ CNOT gates are necessary.

In general case the number of terms in an electronic Hamiltonian is bounded by $O(n^4)$ \cite{Seeley2012}. Since each term can be implemented in $O(n)$ time, one iteration of the phase estimation algorithm would then require $O(n^5)$ quantum gates.
Therefore, these Hamiltonians can be simulated in $O(poly(n))$ time with the truncated Taylor series described in Eq.\eqref{EqExpansionProduct} or with the general circuit in Fig.\ref{FigCircuitAll}} described in this paper.
\section{Writing any $\mH$ as a Sum of Unitaries}
\label{SecIII}
 When $\mH $ is given as a sum of unitary matrices or matrices which can be easily mapped to quantum gates, then one can design the circuit for $\mU\left(t\right)$ by following Eq.\eqref{EqExpansionProduct} where the summation is converted into a product formula or the standard Trotter-Suzuki decomposition. 
 This is the case for molecular Hamiltonians given in the second quantization \cite{Lanyon2010towardsH2,Seeley2012}.
 
 However, if $\mH $ is not given as a sum of simple unitaries\footnote{simple in the sense that the required number of quantum gates for each unitary is polynomial in the number of qubits.}, then the following divide and conquer method can be used to write the Hamiltonian as a sum of unitary matrices and obtain $U_\mH$:
 \begin{enumerate}[label=\roman*)]
\item The matrix is first divided into $2\times2$ submatrices.
\item Then, each submatrix is written as a sum of quantum gates.
\item Using a coefficient matrix $B_\mH$, the Hamiltonian is generated as a part of $U_\mH$.
\end{enumerate}
The details  are given in the following subsections:
\subsection{Division into Submatrices}
First, a given Hamiltonian 
$\mH\in\bf{R}^N$ with  $\left(N=2^n\right)$ is divided into four blocks:
\begin{equation}
	\mH =	\left( \begin{matrix}
	 A_{0}& A_{1}\\ A_{2}&A_{3}
	\end{matrix}\right) = \left( \begin{matrix}
	A_{0}& \\ &A_{3}
	\end{matrix}\right) + \left( \begin{matrix}
	& A_{1}\\ A_{2}&
	\end{matrix}\right).
\end{equation}
Using the vectors in the standard basis, this can be rewritten as:
\begin{equation}
\label{EqHinAs}
\begin{split} 
	\mH  =& \ket{0}\bra{0}\otimes A_0+ \ket{1}\bra{1}\otimes A_3 \\
     & + \left(\ket{0}\bra{1}\otimes A_1 + \ket{1}\bra{0} \otimes A_2\right) \\
    =& \left(\ket{0}\bra{0}\otimes A_0 + \ket{1}\bra{1}\otimes A_3\right)\\ 
    &+ 
     \left(\ket{0}\bra{0}\otimes A_1+ \ket{1}\bra{1}\otimes A_2\right) \left(X\otimes I_{N/2}\right),
\end{split}
\end{equation}
where $I_{N/2}$ describes an $N/2$ dimensional identity matrix
and
 \begin{equation}
X = \left(\begin{matrix}
0&1\\ 1& 0
\end{matrix}\right).
\end{equation}
This division into blocks is recursively continued until each block dimension becomes 2 or circuit representations of the blocks become known.
For instance, after the second recursion step, we have the following (see Appendix \ref{Appendixk2} and \ref{Appendixk3} for the mathematical steps and the third recursion):
\begin{equation}
\label{EqHinAs2}
\begin{split}
\mH = & 
\left( \ket{00}\bra{00}\otimes A_{00} 	
+\ket{01}\bra{01}\otimes A_{03} + 
\ket{10}\bra{10}\otimes A_{30} 	
+\ket{11}\bra{11}\otimes A_{33} \right)
\\ 
 &  +\left(\ket{00}\bra{00}\otimes A_{01}
+\ket{01}\bra{01}\otimes A_{02}
+ \ket{10}\bra{10}\otimes A_{31}
+\ket{11}\bra{11}\otimes A_{32}\right)\left(I\otimes X\otimes I_{N/4}\right)
\\
& +\left(
\ket{00}\bra{00}\otimes A_{10} 	
+\ket{01}\bra{01}\otimes A_{13} +
\ket{10}\bra{10}\otimes A_{20} 	
+\ket{11}\bra{11}\otimes A_{23}\right)\left(X\otimes I\otimes I_{N/4}\right) 
\\
&  +
\left(\ket{00}\bra{00}\otimes A_{11}
+\ket{01}\bra{01}\otimes A_{12}
+\ket{10}\bra{10}\otimes A_{21}
+\ket{11}\bra{11}\otimes A_{22}\right) \left(X\otimes X\otimes I_{N/4}\right).
\end{split}
\end{equation}

In matrix form,
\begin{equation}
\label{EqHinAsMatrix}
\mH =	\left( \begin{matrix}
	\left( \begin{matrix}
	A_{00}& A_{01}\\ A_{02}&A_{03}
	\end{matrix}\right)
		&
		\left( \begin{matrix}
		A_{10}& A_{11}\\ A_{12}&A_{13}
		\end{matrix}\right)
	\\
	\left( \begin{matrix}
	A_{20}& A_{21}\\ A_{22}&A_{23}
	\end{matrix}\right)
		&
		\left( \begin{matrix}
		A_{30}& A_{31}\\ A_{32}&A_{33}
		\end{matrix}\right)
\end{matrix}\right).
\end{equation}
Note that following the subscripts of $A_{[\dots]}$s from left to right one can easily find the matrix elements of any $A_{[\dots]}$.
\subsection{Forming $V_j$s}
By generalizing above steps, at the $k$th recursive step, $\mH$ can be written  more concisely in the following form:
\begin{equation}
\mH=\sum_{j = 0}^{2^k-1}V_j\left(P_j\left(X,I\right)\otimes I_{N/2^k}\right).
\end{equation}
Here, we construct $V_j$s using $A_{[\dots]}$s and  $P_j\left(X,I\right)$ is a permutation matrix constructed by using the tensor product of $X$ and identity matrices, i.e.: 
\begin{equation}
    P_{j}\left(X,I\right) = \bigotimes_{i=0}^{k-1} X^{j_i}, \text{ with }j = \left(j_0\dots j_{k-1}\right)_2.
\end{equation} 
Each $V_j$ describes a multi controlled network: 
In matrix form, $V_j$ is a block diagonal matrix where a group of $A_{[\dots]}$s are tiled on the diagonal. 
\subsubsection{Assigning $A_{[\dots]}$s to $V_j$s at the $k$th step}
Let $w_i$ represent the $i$th word in the set \{``0\dots00", ``0\dots03", \dots, ``3\dots33"\} which includes all possible words from the alphabet \{0,3\}
with the length $k$. 
Then we can define $V_0$ as: 
\begin{equation}
    V_0 = \sum_{i = 0}^{2^k-1} \ket{i}\bra{i}\otimes A_{w_i}, 
\end{equation}
Here, \ket{i}  is the $i$th vector in the standard basis. $V_0$ is obtained from  $A_{[\dots]}$s on the diagonal of the Hamiltonian.
Consider the matrix in Eq.\eqref{EqHinAsMatrix} as an example, then $V_0$ is the block diagonal matrix with $A_{00}, A_{03}, A_{30},$ and $A_{33}$. 

Using $w_i$s in $V_0$ and $P_j\left(X,I\right)$, we will determine the subscripts of  $A_{[\dots]}s$ involved in any $V_j$ by the following rule:
 \begin{rul}
 If there is an $X$ on the $q$th qubit,  the following change is made in the subscripts of $A_{[\dots]}s$ in $V_0$:
    \begin{itemize}
        \item $\left(3 \rightarrow 2\right)$: if the $q$th letter of the subscript is 3, we make it 2.
        \item $\left(0 \rightarrow 1\right)$: if it is 0, we make it 1.
   \end{itemize}
\end{rul}
   You can consider this as an application of NOT gate that switches 3s into 2s and 0s into 1s or vice versa. Based on this rule, one $A_{[\dots]}$ from each row of the Hamiltonian is included in $V_{j}$.

\subsection{Generating Circuit for $U_\mH$ in Eq.\eqref{EqUh}}
If $A_{[\dots]}$s are of dimension $2\times2$, then $2^k = N/2$ and each $V_j$ involves $N/2$ number of $A_{[\dots]}$s. 
Once the involved $A_{[\dots]}$s are determined, they can be mapped to a circuit by using different control bit schemes for each $A_{[\dots]}$. 
Therefore, each $V_j$ describes a multi controlled network.

$A_{[\dots]}$s in general are not unitary. We can write a nonunitary $A_{[\dots]}$ as a sum of two unitary matrices. For instance,
    \begin{equation}
    \label{Equnitaryform}
    A_{[\dots]} = 
    \frac{A_{[\dots]} + i\sqrt{I-A_{[\dots]}^2}}{2} + \frac{A_{[\dots]} - i\sqrt{I-A_{[\dots]}^2}}{2}
    \end{equation} 
This will double the number of $V_j$s. 
Here note that any unitary single qubit gate  can be implemented as a product of three quantum gates and a global phase  \cite{nielsen2002quantum}: 
$e^{i\theta_1} R_z(\theta_2)R_y(\theta_3)R_z(\theta_4)$ with $\theta_1, \dots,\theta_4 \in \bf{R}$. Although this increases the overall number of quantum gates by a factor of 4, it does not impact the number of $V_j$s.

Then, the circuits for $V_jP_j$s are combined in the selection matrix $VP$ by using an ancilla register: 
$VP$ applies the product $V_jP_j$ when the ancilla is in \ket{\bf j} state. 
The matrix form of this operation is as follows: 
\begin{equation}
  VP = \left(\begin{matrix}
  V_0& & &\\
   & V_1P_1 & &\\
   & & \ddots &\\
   & & & V_{N/2-1}P_{N/2-1}
  \end{matrix}\right).
\end{equation}

\begin{figure*}
\begin{center}
\mbox{
\Qcircuit @C=0.15em @R=.3em {
\lstick{\ket{0}}	&	\multigate{1}{B_\mH}	&	\qw	&	\ctrl{2}	&	\ctrlo{1}	&	\ctrlo{1}	&	\ctrlo{1}	&	\ctrlo{1}	&	\ctrlo{1}	&	\ctrlo{1}	&	\ctrlo{1}	&	\ctrlo{1}	&	\ctrl{1}	&	\ctrl{1}	&	\ctrl{1}	&	\ctrl{1}	&	\ctrl{1}	&	\ctrl{1}	&	\ctrl{1}	&	\ctrl{1}	&	\multigate{1}{B^*_\mH}	&	\qw	\\
\lstick{\ket{0}}	&	\ghost{B_\mH}	&	\ctrl{2}	&	\qw	&	\ctrlo{1}	&	\ctrlo{1}	&	\ctrlo{1}	&	\ctrlo{1}	&	\ctrl{1}	&	\ctrl{1}	&	\ctrl{1}	&	\ctrl{1}	&	\ctrlo{1}	&	\ctrlo{1}	&	\ctrlo{1}	&	\ctrlo{1}	&	\ctrl{1}	&	\ctrl{1}	&	\ctrl{1}	&	\ctrl{1}	&	\ghost{B_\mH}	&	\qw	\\
	&	\qw	&	\qw	&	\gate{X}	&	\ctrlo{1}	&	\ctrlo{1}	&	\ctrl{1}	&	\ctrl{1}	&	\ctrlo{1}	&	\ctrlo{1}	&	\ctrl{1}	&	\ctrl{1}	&	\ctrlo{1}	&	\ctrlo{1}	&	\ctrl{1}	&	\ctrl{1}	&	\ctrlo{1}	&	\ctrlo{1}	&	\ctrl{1}	&	\ctrl{1}	&	\qw	&	\qw	\\
	&	\qw	&	\gate{X}	&	\qw	&	\ctrlo{1}	&	\ctrl{1}	&	\ctrlo{1}	&	\ctrl{1}	&	\ctrlo{1}	&	\ctrl{1}	&	\ctrlo{1}	&	\ctrl{1}	&	\ctrlo{1}	&	\ctrl{1}	&	\ctrlo{1}	&	\ctrl{1}	&	\ctrlo{1}	&	\ctrl{1}	&	\ctrlo{1}	&	\ctrl{1}	&	\qw	&	\qw	\\
	&	\qw	&	\qw	&	\qw	&	\gate{A_{00}}	&	\gate{A_{03}}	&	\gate{A_{30}}	&	\gate{A_{33}}	&	\gate{A_{01}}	&	\gate{A_{02}}	&	\gate{A_{31}}	&	\gate{A_{32}}	&	\gate{A_{10}}	&	\gate{A_{13}}	&	\gate{A_{20}}	&	\gate{A_{23}}	&	\gate{A_{11}}	&	\gate{A_{12}}	&	\gate{A_{21}}	&	\gate{A_{22}}	&	\qw	&	\qw	\\
}
}
\end{center}
    \caption{\label{FigCircuitUh}The circuit for $U_\mH$ for any $8\times8$ Hamiltonian matrix: It is assumed that there is no coefficient on any $A_{[\dots]}$. Therefore,  $B_\mH$ can be considered as a tensor product of two Hadamard gates. 
 CNOTs at the beginning of the circuit represents the simplified permutation operations. 
    }
\end{figure*}
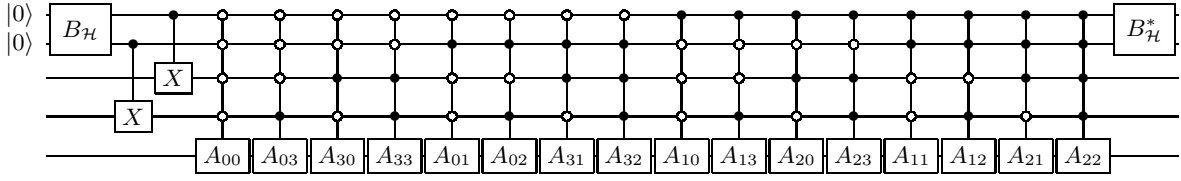

Finally, the circuit for $U_\mH$ implementing the Hamiltonian can be defined as the product 
$\left(B^*_\mH\otimes I\right) VP\left(B_\mH \otimes I\right)$,
 where $B_\mH$  forms a state with the square root of the coefficients: These coefficients are generated by writing nonunitary $A_{[\dots]}$s as a sum of two unitaries. For instance, in Eq.\eqref{Equnitaryform} we have a coefficient 1/2.  This becomes a coefficient to the product $\left(V_jP_j\left(X,I\right)\right)$. 
 $B_\mH$ can be considered as a Householder transformation: An $L$ dimensional Householder matrix requires $O(L)$ number of quantum gates \cite{bullock2005asymptotically,urias2010householder}. 
An example $U_\mH$ for a general $8\times8$ Hamiltonian is presented in Fig.\ref{FigCircuitUh}, where $P_j$s are simplified into two CNOT gates.

Here, note that MATLAB source codes for obtaining $A_{[\dots]}$, $V_j$, and $P_j$ matrices and the circuit can be downloaded from GitHub\footnote{ \url{https://github.com/adaskin/circuitforTaylorseries}}.

\subsection{Gate and Qubit Count}
\subsubsection{Unstructured Dense Matrices}
{
The complexity of $U_\mH$ is determined by the number of $V_j$s.
We can determine the number of qubits and  CNOTs by counting the number of $A_{[\dots]}$s.
In the final step of the recursive division, if the matrix elements are put into group of four elements, then there are $\left(N^2/4\right)$ number of $A_{[\dots]}$s.
If each $A_{[\dots]}$ is written as a sum of two unitary matrices, then the number of unitary gates becomes $\left(N^2/2\right)$. 
As mentioned in the previous subsection, each unitary single gate can be implemented using a product of three rotations. 
Therefore, in total there are three multi controlled-networks with $\left(N^2/2\right)$ number of gates and $\left(2n-1\right)$ number of control qubits.
By following Ref.\cite{mottonen2004circuits}, a network controlled by $\left(2n-1\right)$ qubits can be decomposed into $N^2/2$ CNOTs and $N^2/2$ single gates. Since we have three networks(assuming we have used three single gates for each  unitary gate), then the total number of CNOTs is $3N^2/2$.
The total complexity of a controlled $U_\mH$ is dominated by these networks.
}

{
As a result, a controlled-$U\mH$ in Fig.\ref{FigCircuitAll} can be implemented by using $O(N^2)$ number of quantum gates.
In this case, since the complexity of $B$ and $\Pi$ are negligible in comparison to the controlled-$U_\mH$, the total complexity of the circuit  is also bounded by $O\left(N^2\right)$.
}

\subsubsection{Structured matrices}
{Direct classical methods require $O(N^3)$ computational time (number of floating-point operations) to compute the eigenvalue of a dense matrix.
However, due to discretization and linearization techniques, most of the eigenvalue related-problems deal with structured matrices: that means the description of a matrix depends on less than $N^2$ parameters \cite{fassbender2006structured}.
Many classical algorithms benefit from the structure of a matrix to reduce the computational effort. 
Moreover, in the study of complex many body quantum systems through the random-matrix theory, knowing the structure of the Hamiltonian determines the structure of the random Hamiltonians in the ensemble used to replace the Hamiltonian\cite{guhr1998random}}.

The described divide and conquer method groups the neighboring matrix elements into gates. Any sparsity in the considered matrix may potentially reduce the number of $A_{[\dots]}$s. 
However, when the sparsity of the matrix is structured as in tridiagonal, anti-tridiagonal, and band matrices; the number of terms and so the numbers of qubits and CNOTs used in the circuit are directly affected.
{As an example consider removing the first half of  $A_{[\dots]}$s from the circuit in Fig.\ref{FigCircuitUh}, then we can also remove one of the qubits in the ancilla. This will reduce the gate count by half.
}
{The divide and conquer method does not necessitate to store nonzero elements since the indices are used to determine matrix elements of $A_{[\dots]}$s. Therefore, it can be also used to write sparse matrices in terms of a circuit with $A_{[\dots]}$s. If there are $O(poly(n))$ number of non-zero elements, then this is likely to produce a circuit with $O(poly(n))$ number of $A_{[\dots]}$s.
If any matrix element is accessible in $O(poly(n))$ time, then the construction of the circuit can be done in $O(poly(n))$ time.
}
\subsubsection{Hamiltonian for the Hydrogen Molecule}
Let us consider the $16\times16$ Hamiltonian for the hydrogen molecule given in Eq.\eqref{EqH2Hamiltonian} again. In matrix form, this Hamiltonian  only has 4 non-diagonal elements located on the anti-diagonal part of the Hamiltonian.
If we write the diagonal part and the anti-digonal part as a sum of two unitaries, then the Hamiltonian can be written as sum of 4 number of $\left(V_jP_j\right)$ terms. Since
each $V_j$ involves multi-controlled 8 quantum gates, in total there are 32 quantum gates. Then, $U_\mH$ requires 5 control- and 1 target-qubits.
Since an additional control qubit is necessary for $\mU\left(t\right)$, there are two multi-controlled network, viz. two controlled $U_\mH$s,  with 6 control qubits. 
The decomposition of these networks will constitute $128$ CNOT gates in total.
Here, note that using $\mU\left(t\right)$ in the phase estimation introduces an additional control qubit. Then the required number of CNOTs for each iteration of the phase estimation algorithm is doubled to $\approx256$. 

Here, the hydrogen molecule is given as an example to show how the structured-sparsity may reduce the complexity of the circuit generated through the divide and conquer method. As explained in Sec.\ref{SecIIExampleHamiltonian}, using the Jordan-Wigner\cite{Jordan1928} or Bravyi-Kitaev\cite{BRAVYI2002210} transformations, the molecular Hamiltonians in the second quantization can be easily mapped to a sum of $L$ unitary matrices  (each unitary is a product of Pauli matrices $\sigma_x, \sigma_y, \sigma_z, I$).

\section{Discussion on Adaptive Processes}
\label{Sec4}
 Finding the matrix elements of a Hamiltonian representing the dynamics of a quantum system is a nontrivial task requiring tedious analytical and numerical computations. The circuit can be used to experimentally identify the Hamiltonian dynamic of an unknown large system or estimate the parameters of a quantum channel, where a known state is sent through an unknown state and the measurement is used to estimate the parameters associated with the channel \cite{escher2011general}. 
A similar approach is used also in Ref.\cite{romero2017quantum} to compress molecular Hamiltonians. 
 
\begin{figure}
\begin{center}
\mbox{
\Qcircuit @C=1em @R=.7em {
\lstick{\ket{0}} & \gate{H} &  \ctrl{1}  & \gate{R_z(-\pi/2)}& \gate{H}&\qw \\
\lstick{\ket{\varphi}} & {/} \qw &\gate{U^{2^k}}  &{/} \qw & \qw & \qw &\qw
\\ 
}
}
\end{center}
    \caption{\label{FigIPEA}The $k$th iteration of the iterative phase estimation for adaptive processes: $H$ represents the Hadamard gate and $k \geq 1$.}
\end{figure}
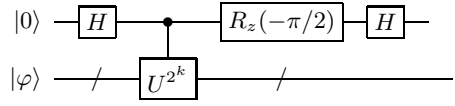

The matrix divided as in Eq.\eqref{EqHinAs2} can be also used to represent a layer of a neural network in matrix form.  In the learning or adaptive processes based on  gradient-descent; when the eigenvalues of an autocorrelation matrix are disparate: i.e. the condition number of the matrix is large, the learning rate or the performance of an adaptive algorithm is hindered in applications \cite{widrow1976stationary}. 
Therefore, in these processes, using the phase estimation as an ingredient for different algorithms may provide better performances.
 The phase estimation algorithm requires two registers to store the eigenvalue and the eigenvector respectively. The size of the first register is determined by the desired accuracy: e.g. for a 32-bit precision, it has 32 qubits.
When the system size is large, 
the iterative phase estimation \cite{Kitaev1996} using only a qubit in the first register is more preferred in the experiments and classical simulations.
However, the iterative version starts the estimation of the bits from the least significant bit (LSB) toward the most significant bit (MSB).
This impedes the employment of the algorithm as a subroutine in various multivariate statistical algorithms
 in which the eigenvalues above or below some threshold are filtered out.
Fig.\ref{FigIPEA} describes the iterative phase estimation algorithm, where the bit values are estimated starting from MSB (see Appendix \ref{AppendixIPEA} for the details of the circuit). This iterative version can be used in adaptive processes to filter out some of the eigenvalues or prepare the ground state of the Hamiltonians.

\section{Conclusion}
In this work, we have introduced a quantum circuit for the simulation of the Hamiltonian dynamic through the Taylor expansion truncated at the third term. The circuit can be used with the Hamiltonians given as a sum of unitary matrices.
Furthermore, we have described a method to write the Hamiltonian as a sum of unitary matrices and generate the equivalent circuit. 
This allows us to use the circuit with the phase estimation algorithm to simulate any Hamiltonian. 
\section{Acknowledgment}
We would like to thank two anonymous reviewers for their help in improving the clarity of the paper and the complexity analysis of the circuit. 

\appendix{
\section{Explicit Steps of the Recursion}
\subsection{Second step $\left(k=2\right)$}
\label{Appendixk2}
First, we plug Eq.\eqref{EqHinAs} in places of  $A_{[\dots]}$s:
\begin{equation}
\begin{split}
\mH  = &\ket{0}\bra{0}\left(
\ket{0}\bra{0}\otimes A_{00} 	
+\ket{1}\bra{1}\otimes A_{03} 
+\left(\ket{0}\bra{0}\otimes A_{01}
+\ket{1}\bra{1}\otimes A_{02}\right)\left(X\otimes I_{N/8}\right)\right)
\\
&+ \ket{1}\bra{1}\left(
\ket{0}\bra{0}\otimes A_{30} 	
+\ket{1}\bra{1}\otimes A_{33} 
+\left(\ket{0}\bra{0}\otimes A_{31}
+\ket{1}\bra{1}\otimes A_{32}\right)\left(X\otimes I_{N/8}\right)\right)
\\
&+  \ket{0}\bra{0}\left(
\ket{0}\bra{0}\otimes A_{10} 	
+\ket{1}\bra{1}\otimes A_{13} 
+\left(\ket{0}\bra{0}\otimes A_{11}
+\ket{1}\bra{1}\otimes A_{12}\right)\left(X\otimes I_{N/8}\right)\right)\left(X\otimes I_{N/4}\right)
\\
&+ \ket{1}\bra{1}\left(
\ket{0}\bra{0}\otimes A_{20} 	
+\ket{1}\bra{1}\otimes A_{23} 
+\left(\ket{0}\bra{0}\otimes A_{21}
+\ket{1}\bra{1}\otimes A_{22}\right)\left(X\otimes I_{N/8}\right)\right)\left(X\otimes I_{N/4}\right) .
\end{split}
\end{equation}
This gives the following:
\begin{equation}
\begin{split}
\mH  = &
\ket{00}\bra{00}\otimes A_{00} 	
+\ket{01}\bra{01}\otimes A_{03} 
+\left(\ket{00}\bra{00}\otimes A_{01}
+\ket{01}\bra{01}\otimes A_{02}\right)
\left(I\otimes X\otimes I_{N/8}\right)
\\
&+ 
\ket{10}\bra{10}\otimes A_{30} 	
+\ket{11}\bra{11}\otimes A_{33} 
+\left(\ket{10}\bra{10}\otimes A_{31}
+\ket{11}\bra{11}\otimes A_{32}\right)
\left(I\otimes X\otimes I_{N/8}\right)
\\
&+ \left(
\ket{00}\bra{00}\otimes A_{10} 	
+\ket{01}\bra{01}\otimes A_{13}\right)
\left(X\otimes I\otimes I_{N/8}\right)
+\left(\ket{00}\bra{00}\otimes A_{11}
+\ket{01}\bra{01}\otimes A_{12}\right)
\left(X\otimes X\otimes I_{N/8}\right)
\\
&+\left(
\ket{10}\bra{10}\otimes A_{20} 	
+\ket{11}\bra{11}\otimes A_{23}\right) 
\left(X\otimes I\otimes I_{N/8}\right)
+\left(\ket{10}\bra{10}\otimes A_{21}
+\ket{11}\bra{11}\otimes A_{22}\right)
\left(X\otimes X\otimes I_{N/8}\right).
\end{split}
\end{equation}
By rewriting this equation, we obtain in Eq.\eqref{EqHinAs2}.
\subsection{Third step $\left(k=3\right)$}
\label{Appendixk3}
Then, in the third recursion, we get the following:
\begin{equation}
\begin{split}
H = &
|0|\otimes A_{000} 	
+|1|\otimes A_{003} 
+ |2|\otimes A_{030} 	
+|3|\otimes A_{033} 
+|4|\otimes A_{300} 	
+|5|\otimes A_{303} 
+|6|\otimes A_{330} 	
+|7|\otimes A_{333}
\\ 
& + 
\left(
|0|\otimes A_{001} 	
+|1|\otimes A_{002} 
+ |2|\otimes A_{031} 	
+|3|\otimes A_{032} 
+|4|\otimes A_{301} 	
+|5|\otimes A_{302} 
+|6|\otimes A_{331} 	
+|7|\otimes A_{332}
\right)\\ 
&\left(I \otimes I\otimes X\otimes I_{N/8}\right)
\\
&+  
\left(
|0|\otimes A_{010} 	
+|1|\otimes A_{013} 
+ |2|\otimes A_{020} 	
+|3|\otimes A_{023} 
+|4|\otimes A_{310} 	
+|5|\otimes A_{313} 
+|6|\otimes A_{320} 	
+|7|\otimes A_{323}
\right)\\ 
&\left(I \otimes X\otimes I\otimes I_{N/8}\right)
\\
&+
\left(
|0|\otimes A_{011} 	
+|1|\otimes A_{012} 
+ |2|\otimes A_{021} 	
+|3|\otimes A_{022} 
+|4|\otimes A_{311} 	
+|5|\otimes A_{312} 
+|6|\otimes A_{321} 	
+|7|\otimes A_{322}
\right)\\ 
&\left(I \otimes X\otimes X\otimes I_{N/8}\right) 
\\
&+ 
\left(
|0|\otimes A_{100} 	
+|1|\otimes A_{103} 
+ |2|\otimes A_{030} 	
+|3|\otimes A_{033} 
+|4|\otimes A_{200} 	
+|5|\otimes A_{203} 
+|6|\otimes A_{230} 	
+|7|\otimes A_{233}
\right)\\ 
&\left(X \otimes I\otimes I\otimes I_{N/8}\right)
\\
&+\left(
|0|\otimes A_{101} 	
+|1|\otimes A_{102} 
+ |2|\otimes A_{031} 	
+|3|\otimes A_{032} 
+|4|\otimes A_{201} 	
+|5|\otimes A_{202} 
+|6|\otimes A_{231} 	
+|7|\otimes A_{232}
\right)\\ 
&\left(X \otimes I\otimes X\otimes I_{N/8}\right)
\\
&+ 
\left(
|0|\otimes A_{110} 	
+|1|\otimes A_{113} 
+ |2|\otimes A_{123} 	
+|3|\otimes A_{123} 
+|4|\otimes A_{210} 	
+|5|\otimes A_{213} 
+|6|\otimes A_{220} 	
+|7|\otimes A_{223}
\right)\\ 
&\left(X \otimes X\otimes I\otimes I_{N/8}\right) 
\\
&+ 
\left(
|0|\otimes A_{111} 	
+|1|\otimes A_{112} 
+ |2|\otimes A_{121} 	
+|3|\otimes A_{122} 
+|4|\otimes A_{211} 	
+|5|\otimes A_{212} 
+|6|\otimes A_{221} 	
+|7|\otimes A_{222}
\right)
\\ &\left(X \otimes X\otimes X\otimes I_{N/8}\right) ,
\end{split}
\end{equation}
where $|i| = \ket{i}\bra{i}$.
This Hamiltonian in matrix form corresponds to the following matrix:
\begin{equation}
\mH =	\left( \begin{matrix}
\left( \begin{matrix}
\left( \begin{matrix}
A_{000}& A_{001}\\ A_{002}&A_{003}
\end{matrix}\right)
& \left( \begin{matrix}
A_{010}& A_{011}\\ A_{012}&A_{013}
\end{matrix}\right)
\\ \left( \begin{matrix}
A_{020}& A_{021}\\ A_{022}&A_{023}
\end{matrix}\right)
&\left( \begin{matrix}
A_{030}& A_{031}\\ A_{032}&A_{033}
\end{matrix}\right)
\end{matrix}\right)
&
\left( \begin{matrix}
\left( \begin{matrix}
A_{100}& A_{101}\\ A_{102}&A_{103}
\end{matrix}\right)
& \left( \begin{matrix}
A_{110}& A_{111}\\ A_{112}&A_{113}
\end{matrix}\right)
\\ \left( \begin{matrix}
A_{120}& A_{121}\\ A_{122}&A_{123}
\end{matrix}\right)
&\left( \begin{matrix}
A_{130}& A_{131}\\ A_{132}&A_{133}
\end{matrix}\right)
\end{matrix}\right)
\\
\left( \begin{matrix}
\left( \begin{matrix}
A_{200}& A_{201}\\ A_{202}&A_{203}
\end{matrix}\right)
& \left( \begin{matrix}
A_{210}& A_{211}\\ A_{212}&A_{213}
\end{matrix}\right)
\\ \left( \begin{matrix}
A_{220}& A_{221}\\ A_{222}&A_{223}
\end{matrix}\right)
&\left( \begin{matrix}
A_{230}& A_{231}\\ A_{232}&A_{233}
\end{matrix}\right)
\end{matrix}\right)
&
\left( \begin{matrix}
\left( \begin{matrix}
A_{300}& A_{301}\\ A_{302}&A_{303}
\end{matrix}\right)
& \left( \begin{matrix}
A_{310}& A_{311}\\ A_{312}&A_{323}
\end{matrix}\right)
\\ \left( \begin{matrix}
A_{320}& A_{321}\\ A_{322}&A_{323}
\end{matrix}\right)
&\left( \begin{matrix}
A_{330}& A_{331}\\ A_{332}&A_{333}
\end{matrix}\right)
\end{matrix}\right)
\end{matrix}\right).
\end{equation}
\section{Iterative Phase Estimation for Adaptive Processes}
\label{AppendixIPEA}
\begin{figure}
    \includegraphics[width=3in]{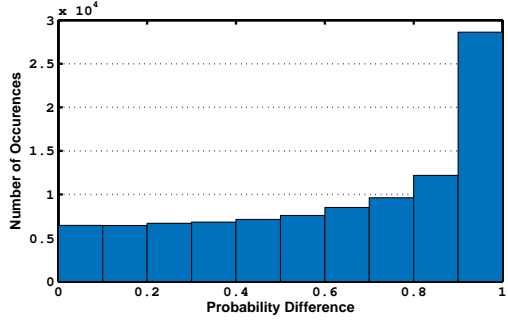}
    \caption{\label{FigHist}The histogram of the probability differences in the outputs of the iterative phase estimation for 5000 random matrices. The algorithm is iterated 20 times for each matrix. Here, the probability to see the probability difference less than 0.1 is 0.0644 and greater than 0.9 is 0.2862.}
\end{figure}

The circuit in Fig.\ref{FigIPEA} works as follows:
\begin{itemize}
\item Assume that the circuit $U= e^{i\mH\pi}$ and its eigenvector $\ket{\varphi}$ are given. 
\item In the first iteration $k$ is set to $1$. After the controlled-$U$, then the following state is obtained (Normalization constants are omitted for simplicity.):
\begin{equation}
\ket{\psi_1} =\ket{0}\ket{\varphi}+e^{i\pi\left(\phi_1.\phi_2\dots\right)_2}\ket{1}\ket{\varphi}. 
\end{equation} 
Here, $\left(\phi_1.\phi_2\dots\right)_2$ represents the binary form of the phase multiplied by 2: $2\phi$.
\item To make the real part of the value $e^{i\pi\left(\phi_1.\phi_2\dots\right)_2}$  negative when $\phi_1 =1$ and positive when $\phi_1 =0$, we apply a rotation  $R_z\left(-\pi/2\right)$:
\begin{equation}
R_z\left(-\frac{\pi}{2}\right) = 
\left(\begin{matrix} 1&0\\
0& e^{-i\pi /2}
\end{matrix}\right).
\end{equation}
Note that in general when rotation gate about the z-axis is defined, the angle is divided by 2, which is neglected here.
After this gate, we have:
\begin{equation}
\ket{\psi_2} =\ket{0}\ket{\varphi}+e^{i\pi\left(\phi_1.\phi_2\dots\right)_2-i\pi/2}\ket{1}\ket{\varphi}. 
\end{equation} 

\item After applying the second Hadamard gate, the final state becomes the following:
\begin{equation}
\begin{split}
\ket{\psi_3} = & \left(1+\cos\left(\alpha\right)+i\sin\left(\alpha\right)\right)\ket{0}\ket{\varphi} \\
&+ \left(1-\cos\left(\alpha\right)-i\sin\left(\alpha\right)\right)\ket{1})\ket{\varphi},
\end{split}
\end{equation}
where $\alpha =  \pi\left(\phi_1.\phi_2\dots\right)_2-\pi/2$. 
The probability difference between \ket0 and \ket1 is determined by the value of $\cos(\alpha)$:
\begin{itemize}
\item If $\phi_1 = 0$, then $\alpha\in \left[-\frac{\pi}{2},\frac{\pi}{2}\right]$ and $\cos(\alpha)\geq 0$. Hence, the probability of \ket0 is higher than \ket1.
\item If $\phi_1 = 1$, then $\alpha\in \left[\frac{\pi}{2},\frac{3\pi}{2}\right]$ and $\cos(\alpha)\leq 0$.
And so, the probability of \ket1 is higher than \ket0. 
\end{itemize} 
\item In the second iteration; $k=2$ and if $\phi_1$ is 1, then we have  $(10)_2 + (\phi_2.\phi3\dots)_2$.  Since $e^{i(10)_2\pi}=1$, the phase is in the form $(\phi_2.\phi3\dots)_2$. 
\end{itemize}
The probability difference between \ket0 and \ket1 in the output may be as high as 1 or in rare cases  is equal to 0.
For random 5000 matrices, the distribution of the probability differences is drawn Fig.\ref{FigHist}: The probability differences are obtained by iterating the phase estimation algorithm 20 times for each matrix. 
As shown in the figure, the probability to see  the difference  less than 0.1 is  around 0.06.

}

\end{document}